\documentclass[sigconf,authorversion]{acmart}

\usepackage[normalem]{ulem}
\AtBeginDocument{%
  \providecommand\BibTeX{{%
    \normalfont B\kern-0.5em{\scshape i\kern-0.25em b}\kern-0.8em\TeX}}}

\copyrightyear{2024} 
\acmYear{2024} 
\setcopyright{rightsretained} 
\acmConference[WSDM '24]{Proceedings of the 17th ACM International Conference on Web Search and Data Mining}{March 4--8, 2024}{Merida, Mexico}
\acmBooktitle{Proceedings of the 17th ACM International Conference on Web Search and Data Mining (WSDM '24), March 4--8, 2024, Merida, Mexico}\acmDOI{10.1145/3616855.3635727}
\acmISBN{979-8-4007-0371-3/24/03}

\usepackage{booktabs} 
\usepackage{color}

\begin{document}

\title{Grounded and Transparent Response Generation for Conversational Information-Seeking Systems} 

\author{Weronika Łajewska}
\affiliation{%
  \institution{University of Stavanger}
  \city{Stavanger}
  \country{Norway}
}
\email{weronika.lajewska@uis.no}
\orcid{0000-0003-2765-2394}

\begin{abstract}
While previous conversational information-seeking (CIS) research has focused on passage retrieval, reranking, and query rewriting, the challenge of synthesizing retrieved information into coherent responses remains. The proposed research delves into the intricacies of response generation in CIS systems. Open-ended information-seeking dialogues introduce multiple challenges that may lead to potential pitfalls in system responses. The study focuses on generating responses grounded in the retrieved passages and being transparent about the system's limitations. Specific research questions revolve around obtaining confidence-enriched information nuggets, automatic detection of incomplete or incorrect responses, generating responses communicating the system's limitations, and evaluating enhanced responses. By addressing these research tasks the study aspires to contribute to the advancement of conversational response generation, fostering more trustworthy interactions in CIS dialogues, and paving the way for grounded and transparent systems to meet users’ needs in an information-driven world.
\end{abstract}



\maketitle

\section{Introduction}

The ever-growing reliance on digital information demands the presence of transparent and trustworthy search systems in our daily interactions. A large fraction of research on conversational information-seeking (CIS) systems to date has focused on the problem of passage retrieval~\citep{Luan:2021:TACL}, reranking~\citep{Pradeep:2021:arXiva}, and query rewriting~\citep{Vakulenko:2021:WSDM}. However, identifying the top relevant passages is only an intermediate step. Ultimately, the information from these passages would need to be synthesized into a single response in the process of \emph{conversational response generation}~\citep{Ren:2021:TOIS}. The response returned by the system should encapsulate the most relevant pieces of information in an easily consumable unit~\citep{Culpepper:2018:SIGIRa}. 

Generative language models have been widely adopted for response generation~\citep{Zhang:2020:ACL, Lewis:2020:NIPS}; however, in the realm of open-ended infor\-mation-seeking dialogues, the assumption that a user's query can be definitively answered by simply summarizing information from top retrieved passages falls short of reality. System responses are susceptible to various limitations, such as the failure to find a response which may result in hallucinations~\citep{Ji:2023:ACM}, providing a biased response only partially answering the question~\citep{Gao:2020:Inf}, or even presenting content with factual errors~\citep{Tang:2022:NAACL-HLT}. Consequently, relying solely on summarizing relevant information may lead to providing users with biased, incomplete, or, worse, incorrect responses~\citep{Tang:2022:NAACL-HLT}.

Recognizing that users are responsible for judging the completeness, credibility, and accuracy of information provided by the system, there arises a crucial need to equip them with the necessary tools for objective assessment. Two key elements come into play for this purpose: (1) disclosing the system's confidence in the response and (2) providing transparency on system limitations. 
By showing how confident the system is in its response, users can assess the reliability of the information provided~\citep{Lu:2021:CHI, Rechkemmer:2022:CHI, Shani:2013:Jb}. This helps differentiate between well-supported responses grounded in specific facts and responses with more uncertainty.
Equally important is disclosing the limitations inherent in the response generation process. Users must be informed about potential factors that could have contributed to inaccurate responses, such as unanswerability~\citep{Sulem:2022:NAACL-HLT, Choi:2018:EMNLP, Rajpurkar:2018:ACL},
lack of viewpoint diversification~\citep{Gao:2020:Inf, Draws:2021:SIGIR},
or queries with impossible conditions~\citep{Lee:2020:LREC}. 
Understanding these limitations empowers users to interpret responses more critically and aids in their decision-making process. 

\begin{figure*}
    \centering
    \includegraphics[width=1.0\textwidth]{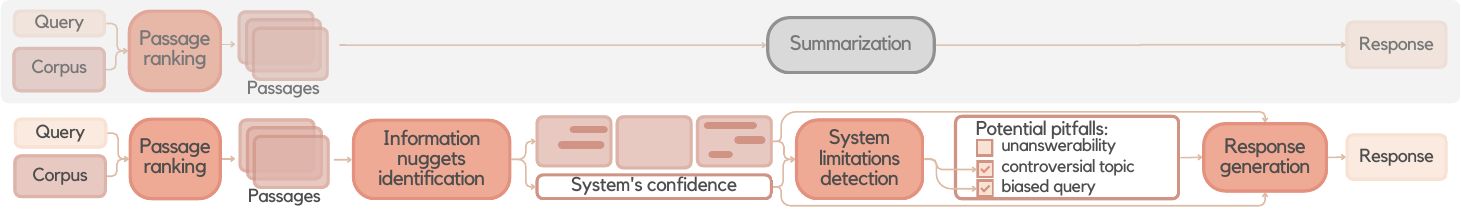}
    \vspace*{-0.45cm}
    \caption{Overview of our approach to conversational response generation in comparison with the standard CIS pipeline.}
    \label{fig:system-architecture}
    \vspace{-0.18cm}
\end{figure*}

\textbf{RQ1: How to detect factors contributing to incorrect, incomplete, or biased responses?}
One of the primary concerns in response generation is how to obtain information nuggets~\citep{Ren:2021:TOIS} (RQ1.1) defined as \emph{minimal, atomic units of relevant information}~\citep{Pavlu:2012:WSDM} that contain key pieces of information necessary to answer the user's question. In open-ended information-seeking dialogues, the system's ability to pinpoint crucial details is essential for fact-grounded response generation. We aim to develop methods and techniques to extract these confidence-enriched information nuggets from the top retrieved passages, building foundations for more factual system responses.
Given that CIS systems are susceptible to limitations, it raises a question about factors that lead to incomplete, biased, or incorrect responses (RQ1.2).  
We propose automated techniques to detect underlying factors leading to flawed responses by analyzing queries and information nuggets. This approach aims to unveil patterns indicating response pitfalls, enhancing the system's ability to identify problematic responses generated from information nuggets.

\textbf{RQ2: How to generate responses transparent about the system's confidence and limitations?}
Providing users with grounded and transparent responses is crucial for fostering trust and empowering users to objectively assess the presented information. We investigate how to generate responses that not only synthesize the requested information but also are grounded in the specific facts identified in the passages and articulate the system's confidence and limitations. Designing informative responses that transparently present potential pitfalls will promote a more informed user experience and help manage user expectations~\citep{Radlinski:2017:CHIIR, Azzopardi:2018:CAIR}. 
The accuracy of generated responses and effectiveness in communicating the system's confidence and limitations can be evaluated via user studies. By exposing users to different variants of the responses, we aim to evaluate our approach to enhancing the transparency of CIS systems. 
Additionally, we aim to propose measures for automatic evaluation of the generated responses quantitatively, in terms of response completeness using the identified information nuggets.

By addressing these research tasks, our study 
aspires to bridge the gap between system-generated responses and user comprehension, fostering a more trustworthy interaction in CIS dialogues.
\section{Methodology}
\label{sec:method}

Our proposed response generation module, in contrast to the summarization of top retrieved passages, identifies information nuggets and aggregates them to the response~\citep{Ren:2021:TOIS}. 
Additionally, the response (see~Fig.~\ref{fig:system-architecture}) is enriched with the information about system's confidence and the potential factors that could have contributed to flaws.

\subsection{Passage Ranking}

The main component of the CIS pipeline is passage ranking, retrieving the most relevant passages from the corpus. This study focuses on downstream processing following passage retrieval, and not on enhancing the ranking methods. Thus, we employ the top-performing 2021 TREC Conversational Assistance track submission reproduced in our prior work~\citep{Lajewska:2023:ECIR}, serving as a strong baseline.

\subsection{Information Nuggets Identification}

To address RQ1.1, which concerns obtaining confidence-enriched information nuggets, we propose methods for extracting key pieces of information from passages to ensure fact-grounded response generation. 
In recent work, we have created the CAsT-snippets dataset~\citep{Lajewska:2023:}, enriching the TREC CAsT 2020~\citep{Dalton:2020:TREC} and 2022~\citep{Owoicho:2022:TREC} datasets with snippet-level answer annotations. To ensure data quality, we extensively investigated effective task designs and trade-offs for snippet annotations through crowdsourcing, considering various interfaces and worker qualifications. The primary challenge in gathering snippet annotations centered around the quality control process, which couldn't be automated due to the inherent nature of the task. 
In comparison to other related datasets (SaaC~\citep{Ren:2021:TOIS}, QuaC~\citep{Choi:2018:EMNLP}), CAsT-snippets not only contains more snippets annotated for each input text, but they are also longer on average.
In future work, we aim to focus on the automatic identification of information nuggets in passages and determining the system's confidence in the selected snippets. Information nugget identification is not a binary decision but a more granular task, as snippets can vary in relevance and complexity. They may contain exact facts fully answering the question, additional enriching details, or only address some aspects of the question. This granular nature of \emph{answerability} presents a major challenge and plays a crucial role in incorporating the system's confidence score in the generated response.

\subsection{Detecting System Limitations}

To address RQ1.2, which concerns the identification of potential system limitations while generating the response from identified information nuggets, we propose to detect factors that may lead to incomplete, or incorrect responses. 
As we seek to detect these limitations from the system's perspective, we lack external ground truth for comparing different systems, particularly for evaluating the completeness of responses. In recent work, we proposed a mechanism for detecting unanswerable questions where the correct answer is not present in the corpus or cannot be retrieved. 
We develop a method that employs a sentence-level classifier to detect if the answer is present, then aggregates these predictions on the passage level, and eventually across the top-ranked passages to arrive at a final answerability estimate. For training and evaluation, we extend the CAsT-snippets dataset with answerability labels on the sentence, passage, and ranking levels.
By assessing if a question can be at least partially answered based on the information contained in the top-ranked passages, we can mitigate the risk of generating responses from irrelevant or non-existent answers, reducing the occurrence of hallucinations~\citep{Ji:2023:ACM}. 
In our future work, we aspire to extend our detection capabilities to encompass other limitations that influence the final step of response generation. These limitations include but are not limited to, lack of viewpoint diversification when addressing controversial topics~\citep{Draws:2021:SIGIR}, partial unanswerability, temporal considerations
~\citep{Campos:2015:ACM}, biased queries~\citep{Azzopardi:2021:CHIIR}, the subjectivity of the source text, and the lack of 
expert/background knowledge required to infer the answer.
\subsection{Revealing System Confidence and Limitations}

RQ2 concerns the generation of grounded responses that reveal system confidence and limitations.
In recent work, we ran two crowdsourcing experiments investigating user perceptions of problems of unanswerable questions and incomplete responses in CIS systems. The results show that users find it easier to detect issues related to viewpoint diversity and response bias compared to factual errors and source validity. Our findings imply that simple source attribution is insufficient for effective system interaction, suggesting the need for explicit communication of potential inaccuracies to enhance users' assessments of the presented information.
Providing users with transparent responses that acknowledge the system's limitations is paramount for fostering trust and empowering users to make informed judgments~\citep{Radlinski:2017:CHIIR, Azzopardi:2018:CAIR}. We aim to generate responses that (1) synthesize the requested information, (2) ground it in specific facts identified in the passages, (3) articulate the system's confidence, and (4) reveal the system's limitations. By openly showing users the system's limitations, we aim to encourage a more critical examination of the responses, prompting them to verify the provided information. 
We consider diverse methods for conveying this information to users, including natural language incorporation~\citep{Rechkemmer:2022:CHI}, UI elements~\citep{Lu:2021:CHI}, and a granular confidence scale~\citep{Shani:2013:Jb}, drawing from prior work in recommender systems that can be adapted to the field of conversational response generation.

Evaluating generated responses in CIS systems in terms of completeness is challenging, as the system cannot be fully aware of what it does not know, even with identified information nuggets~\citep{Chen:2021:ACL-IJCNLP}. However, we recognize that snippet-level answer annotation can be a crucial first step toward automatically evaluating responses quantitatively, based on the relevant information nuggets included~\citep{Pavlu:2012:WSDM}.
The recent edition of TREC CAsT proposes to evaluate summarized passages based on relevance, naturalness, and conciseness~\citep{Owoicho:2022:TREC}. We aim to extend the evaluation of generated responses from the perspective of transparency and grounding.
The question about other aspects of response generation that should be considered, in addition to transparency and grounding, remains.
User feedback from our crowdsourcing experiments related to challenges in CIS response generation draws attention to the credibility of the sources, as well as the completeness, usefulness, and subjectivity of provided information that impact users' overall satisfaction. 

\section{Research issues}
\label{sec:research_issues}

In Section~\ref{sec:method}, we established a set of simplifying assumptions for our proposed approaches. Now, we highlight the open questions we have identified, seeking assistance in prioritizing them due to their expansive nature and the time constraints of the Ph.D. study. 
We seek for guidance on the following issues related to our research questions: (RQ1.1) How can the system articulate its confidence in the identified snippets without information about the scope of the complete response? (RQ1.2) Should we focus on identifying more limitations or delve deeper into specific ones? (RQ1.2) Can limitation detection be addressed holistically? (RQ2) Are the attempts to showcase system limitations through additional user interface elements justifiable, or does the superiority still lie in the simplicity of generating natural language responses? 

\bibliographystyle{ACM-Reference-Format}
\bibliography{wsdm2024-dc_wl.bib}

\end{document}